\documentclass[reprint,secnumarabic,amssymb,amsmath,aip,cha,groupedaddress,frontmatterverbose]{revtex4-1}
\usepackage{docs,graphicx,bm}

\begin{document}

\title{Ratcheted diffusion transport through crowded nanochannels}

\author{Anna Lappala, Alessio Zaccone, Eugene M. Terentjev}
\email{emt1000@cam.ac.uk}
\affiliation{Cavendish Laboratory, University of Cambridge, JJ Thomson
Avenue, Cambridge CB3 0HE, U.K.}

\begin{abstract}
The problem of transport through nanochannels is one of the major questions in
cell biology, with a wide range of applications. Brownian ratchets are
fundamental in various biochemical processes, and are roughly divided into two
categories: active (usually ATP-powered) molecular motors and passive constructions 
with a directional bias, where the transport is driven by thermal motion. 
In this paper we discuss the latter process, of spontaneous translocation of 
molecules (Brownian particles) by ratcheted diffusion with no external energy input: 
a problem relevant for protein translocation along bacterial flagella or injectosome complex, or DNA translocation by bacteriophages. We use molecular dynamics simulations and statistical theory to identify two regimes of transport: at low rate of particles injection into the channel the process is controlled by the individual diffusion towards the open end (the first passage problem), while at a higher rate of injection the crowded regime sets in. In this regime the particle density in the channel reaches a constant saturation level and the resistance force increases substantially, due to the osmotic pressure build-up. To achieve a steady-state transport, the apparatus that injects new particles into a crowded channel has to operate with an increasing power consumption, proportional to the length of the channel and the required rate of transport. The analysis of resistance force, and accordingly -- the power required to inject the particles into a crowded channel to oversome its clogging, is also relevant for many microfluidics applications.
\end{abstract}

\maketitle

Ratcheted diffusion can be described as directionally biased random
thermal motion. Physically, this process is often described as a transport
phenomenon in spatially periodic systems, a typical example of which is a
``saw-tooth'' potential that is periodic in space and has a broken
forward-backward symmetry, however, ratchet potentials exist in various forms
in terms of symmetry and periodic behavior.\cite{reimann_2002}
The function of biologically relevant ``motor" enzymes, or molecular
motors, is based on non-equilibrium ratcheted diffusion, where the energy is 
continuously recycled in the system -- typically by the ATP hydrolysis. Well studied 
classes of these
molecular motors are myosins, kinesins and dyneins \cite{reimann_2002}.
The physical mechanism producing the powered
linear motion along the filament with asymmetric (saw-tooth)
potential, energized by the ATP hydrolysis, is well understood after the
classical work of Prost et al. \cite{prost1,prost2}  Myosins
move along actin filaments, and these proteins are responsible for muscle
contraction \cite{berg_2007}. Kinesin and dynein motor proteins move along
microtubules, either towards the elongating/polymerising $+$ end (which has a
high concentration of so called $\beta$-subunits) or the $-$ end (which has a
high concentration of $\alpha$-subunits). Dyneins have various additional
functions which include facilitating the movement of cilia and flagella.\cite{berg_2007}

Our interest is different:
we examine the equilibrium situation when the directed transport is achieved spontaneously,
essentially, energized by the Brownian thermal motion and directed by restricting the
diffusion in the ``wrong'' direction. Examples of this equilibrium diffusive transport
is the translocation of cargo across membranes through nanopores
or along nanochannels.\cite{hille2001} Essentially, the directed Brownian motion is
achieved by simply not allowing the cargo to
travel back to the side of the channel where the process of translocation
started. To achieve this, the cargo has to block the channel width such that no overtaking is possible in a single-file (one-dimensional) motion.\cite{bechinger2000} In this case the potential profile along the channel is irrelevant,
as the equilibrium detailed balance would not produce any directed motion without an external force
(in spite of any symmetry breaking). A variety of chemical asymmetries may exist inside the channel to affect the movement of the cargo.\cite{simon_1992} However, the effect of the channel potential is merely to modify the effective diffusion constant, which is again a well-studied problem.\cite{lubensky_1999} Biological examples of transport assumed to rely on spontaneous ratcheted diffusion are found in various biological systems, such as:

\noindent $1)$ flagella -- hair-like protrusions from the cell body of specific
pro- and eukaryotic cells \cite{wang_2005} grow by transporting
flagellin protein subunits from the body of bacteria through a 2nm wide channel
over distances of 10-20 $\mu$m, at a constant rate; \cite{berg_2012}

\noindent $2)$ the injectosome complex (type \MakeUppercase{\romannumeral
3}~secretion system) -- a needle-like protein appendage that is found in
gram-negative bacteria, the function of which is to secrete proteins into
eukaryotic host cells to facilitate the process of infection
\cite{salmond_1993};

\noindent$3)$ the general secretory (Sec) pathway, in which the translocase
protein complex machinery provides a translocation pathway for proteins from
the cytosol across the cytoplasmic membrane in bacteria;\cite{pugsley_1993}

\noindent$4)$ bacteriophage (bacterial virus) DNA injection into host bacterial
cells through the tail of the phage -- a good example of this process is the T4
phage mechanism of injection.\cite{furukawa_1983}

Independently, ratcheted transport in Brownian systems is of great practical relevance
for applications in lab-on-chip microfluidic devices.\cite{weitz_2008} Typically, if crowding conditions of drops/bubbles/particles are achieved
at the inlet of the microfluidic channel, the resistance to the transport develops, which eventually lead to microchannel clogging.\cite{whitesides} The latter
is a practically important phenomenon which limits the performance of microfluidic devices and of industrial micromixers.\cite{weitz_2006,ottino}

One particular biological system that we shall frequently refer to is
the bacterial flagellum, which  is
a tube with a 2nm wide inner channel and the walls formed by crystallized flagellin proteins. In
order for the flagellum to grow, the protein subunits
have to be translocated to the distal (growing) end of the flagellum
through this channel, which is too narrow to accommodate folded proteins.
Hence, subunits are transported in their {unfolded} state, after they are
inserted into the channel by the {export apparatus}.\cite{ibuki_2011}
Flagellar export apparatus is a multicomponent complex, energized by ATP hydrolysis and
achieving the successive unfolding of flagellin chains and insertion of them
into the channel,\cite{minamino_2008,paul_2008,galan_2008} thus providing a bias
of the subsequent 1D Brownian motion, which motivates the work presented in this paper.

Very recently there has been an attempt on studying the diffusion of flagellin subunits along 
this channel,\cite{berg2013} where many of the details and parameters of this biological 
system are carefully presented. The shortcoming of this basic simulation is that it does not 
take into account interactions (essentially -- repulsion on contact) between travelling particles, 
which then fails to describe the crowding regime.
Here we show molecular-dynamics simulation results for translocation of Brownian particles
in nanochannels under conditions at the inlet, which mimic crowded cellular environments.
The effect of molecular crowding at the channel inlet is twofold: on the one hand it acts as a net drift speeding up the translocation near the open outlet, but on the other hand it also causes an increasing resistance force, slowing down the transport near the channel inlet. The range of channel length where crowding contributes a net resistance force increases with the total channel length. Remarkably, this effective resistance force is an emergent phenomenon which in 1D depends entirely on the cooperative multi-body dynamics of particles colliding in the channel.
Collective diffusion combined with a third-order virial expansion provides a simple theoretical description
of the density profile inside the channel in agreement with the simulation data.
The transition from classical free diffusion to the crowding regime is controlled by the competition between the particle injection rate and the mean first passage time of free diffusion. This fact is crucial in order to assess the onset of crowding in both biological and engineering systems.

\section*{Free single-file diffusion}

The ratcheting, or \emph{rectification} of the Brownian motion along the narrow
channel is achieved by continuously inserting new particles from one end,
thus restricting the backward motion -- while the channel remains open at the
distal end. The diffusion process as described in this paper raises three main
issues: $1)$ how high is the resistance force, if the particles are inserted at
a sufficiently high rate? $2)$ How do the two characteristic times, the mean
first passage time $\tau_{\textrm{diff}}$ and the rate of insertion
$1/T_{\textrm{in}}$ compete? $3)$ What is the crossover between the stochastic
diffusion dominated regime (when $T_{\textrm{in}}\gg\tau_{\textrm{diff}}$) and
the dynamically restricted, resistance-dominated regime in a crowded channel?

\begin{figure}[t]
\begin{center}
  \includegraphics[width=0.35\textwidth]{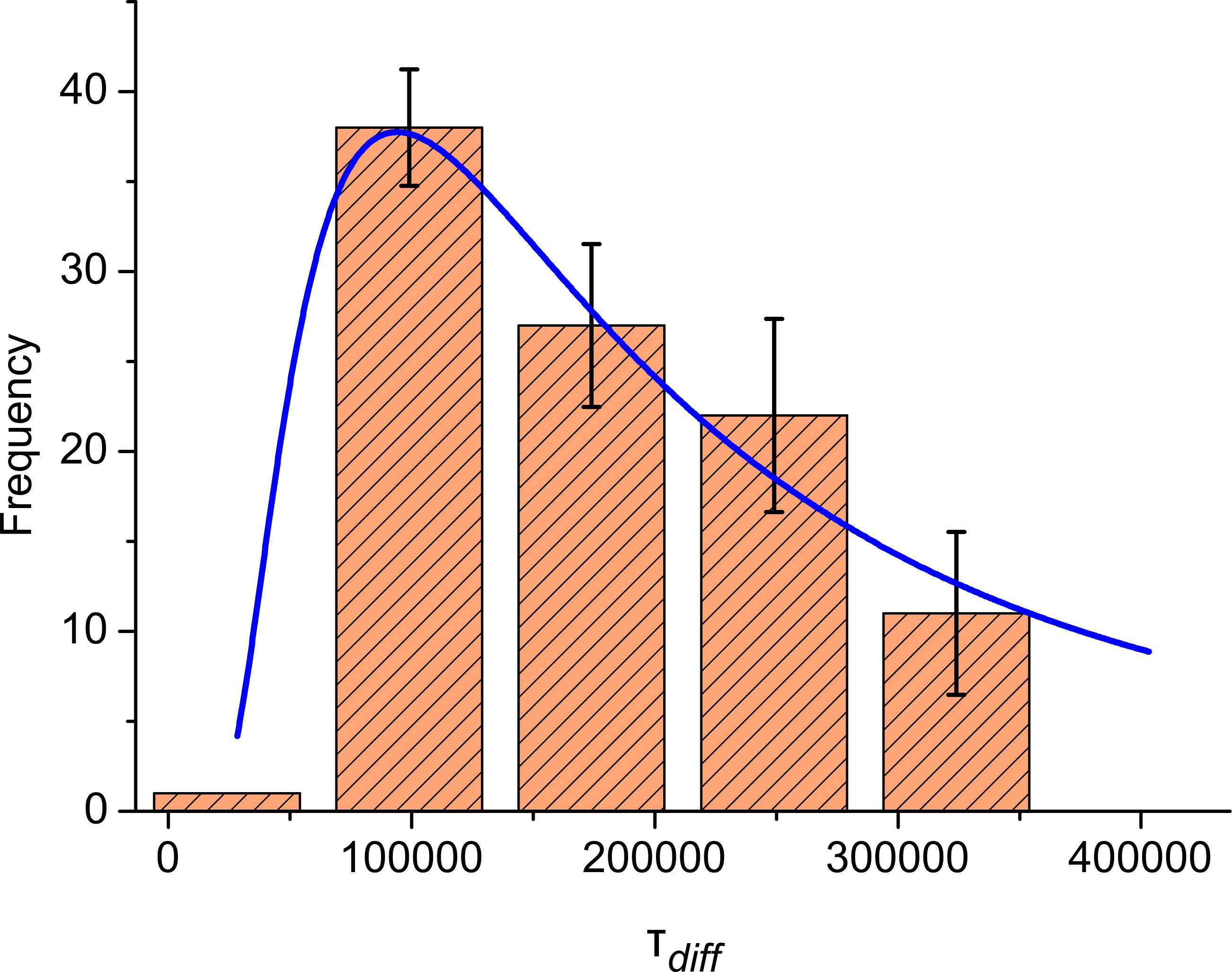}
\caption{A histogram representing the distribution profile of individual diffusion times,
$\tau_{\textrm{diff}}$, along the channel of length $L=60 \sigma$ ($\sigma$ being the size of particle). The time is measures in [kts]: 1000 of simulation time steps.
The solid line represents the fit to the probability distribution
discussed in the text, eq.\eqref{Tdiff}, with only the normalization factor and the diffusion
constant $D$ as fitting parameters.  }\label{fig:tau}
\end{center}
\end{figure}

In order to introduce the problem, we first look at the motion of a single
particle inside a channel blocked at one end, in order to verify the properties
of generalized diffusion, fig.\,\ref{fig:tau}.
Since the particle essentially
disappears as soon as it reaches the open end, a distance $L$ away, this is a
problem of first passage with a reflecting boundary condition at $x=0$ and an
absorbing condition at $x=L$.
The probability distribution of first passage times
is a classical result obtained by
the method of images due to Lord Kelvin;\cite{feller_1971} it takes the
form (see ``Methods'' for detail):
\begin{equation}
f(t ) = \textrm{const} \cdot \frac{L}{2 \sqrt{\pi D t^3}} \left(e^{-L^{2}/4D
t}-e^{-L^{2}/Dt} \right).  \label{Tdiff}
\end{equation}
This probability distribution is used to fit the histogram of simulation results for the frequency of diffusion time of a single particle along the channel, obtained by the Brownian dynamics simulation, fig.\,\ref{fig:tau}.
In particular, the free diffusion constant in our simulation is thus measured to be $D = 0.0065$ in the reduced simulation units of [$\sigma^2/\tau$]. An example analysis in ``Methods'' shows that this corresponds to a value $D =6.5\cdot 10^{-12}\textrm{m}^2/\textrm{s}$ for 1-micron size polystyrene colloids in water.

Unlike earlier work,\cite{reimann_2002,lubensky_1999} we do not consider designed periodic potentials in the
channel for simplicity, and due to the fact that this is not necessary in order
to find $D$, the diffusion constant. The important result, reproduced frequently in the literature, is that the effective diffusion constant for traveling in a channel with a potential $V(x)$ along its length is related to the bare diffusion constant (with no potential) as~\cite{lifson_1962}
\begin{equation}
D= \frac{D_0}{\int_0^L \frac{dx_1}{L} e^{-V(x_1)/k_BT}  \int_0^L \frac{dx_2}{L} e^{V(x_2)/k_BT}  }. \label{Deff}
\end{equation}
The product of two integrals in the denominator is always greater than unity (this is demonstrated by applying the Cauchy-Schwartz inequality,\cite{lifson_1962} see also~\cite{zwanzig}) and therefore, for any potential $V(x)$ on the channel walls the diffusion is hindered. For our purposes of studying the effects of crowding in the channel, we may simply take a given fixed value for the effective diffusion constant, $D$, obtained by the fitting of single-particle diffusion data as in fig.\,\ref{fig:tau}, and proceed to examine the case when many particles restrict their motion by crowding.

\section*{Crossover to crowded regime}

Having determined the average time of individual particle diffusion to the far end of the channel, $\tau_{\textrm{diff}} = \int t \cdot f(t) dt = L^2/D$, we can now simulate the process with particles injected at $x=0$ with increasingly short time intervals $T_{\textrm{in}}$ (i.e at an increasing rate, or the particle flux $J_\textrm{in} = 1/T_{\textrm{in}}$). As explained in ``Methods'' below, our simulation algorithm only allows the insertion if a sufficient room (of the particle size $\sigma$) is available at $x=0$; if the space is occupied, the insertion is missed and is attempted after the next $T_{\textrm{in}}$ interval. At low rate of insertion this omission never happens, but at $T_{\textrm{in}} \ll \tau_{\textrm{diff}}$ this becomes a limiting factor. Figure~2 represents the results in the form of the cumulative number of particles having passed through the channel: $N(t)$ is essentially the sequential number of a particle leaving the channel at a time $t$. From the plot it is clear that at sufficiently long times of simulation the progression of particles is linear on average. One may interpret this as a constant rate of delivery of cargo along the channel, however, we shall discuss below the constraints on the required power and the force exerted by the insertion apparatus.

\begin{figure}[h]
\begin{center}
  \includegraphics[width=0.37\textwidth]{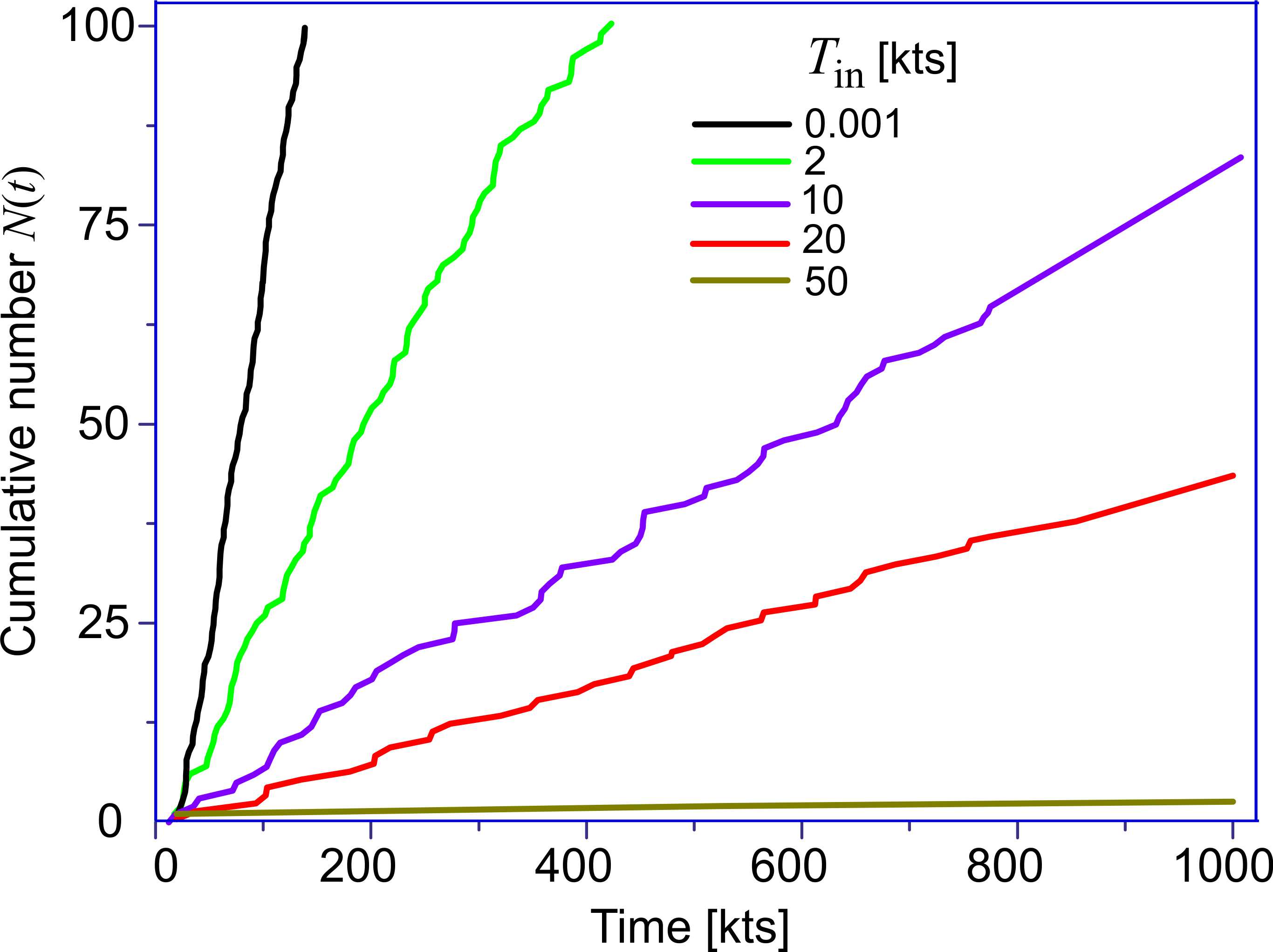}
\caption{A graph of a cumulative number of particles having passed through the channel of $L=30\sigma$ against simulation time-step (measured in [kts]: 1000's of time steps, as in fig.\,\ref{fig:tau}). The labeled values of $T_{\textrm{in}}$, also in [kts], represents the set interval of an attempted insertion at $x=0$. The large scale of this cumulative plot makes obvious the linearity of $N(t)$, but masks what happens at shorter times -- when the insertion location may be occupied by other particles; fig.\,\ref{fig:flux}  below illustrates the effect of crowding restriction.}\label{fig:cumulative}
\end{center}
\end{figure}

Figure~3 plots the average slope of the data in fig.\,\ref{fig:cumulative}
on the double-logarithmic scale to allow examination of the two key findings. We choose to plot the inverse flux, $dt/dN(t)$ against $T_{\textrm{in}}$, to have a more clear sense of the events. The plot also shows the results for three different channel lengths (measured in the units of particle size $\sigma$), as labeled on the figure. In essence, every curve in fig.\,\ref{fig:cumulative} gives a single point in this graph, for a given $L$ and $T_{\textrm{in}}$. Two regimes are evident: when particles are inserted at a sufficiently low rate, the insertion rate is simply equal to the constant flux $J_0$ through the channel, with a small correcting factor reflecting the channel length. However, when the attempted rate of insertion becomes high, the rate of diffusive transport through the channel saturates at a constant maximum value that depends on the channel length and is determined by the increasing resistance for the motion of the single file of particles packed in the channel.

\begin{figure}[h]
\begin{center}
  \includegraphics[width=0.36\textwidth]{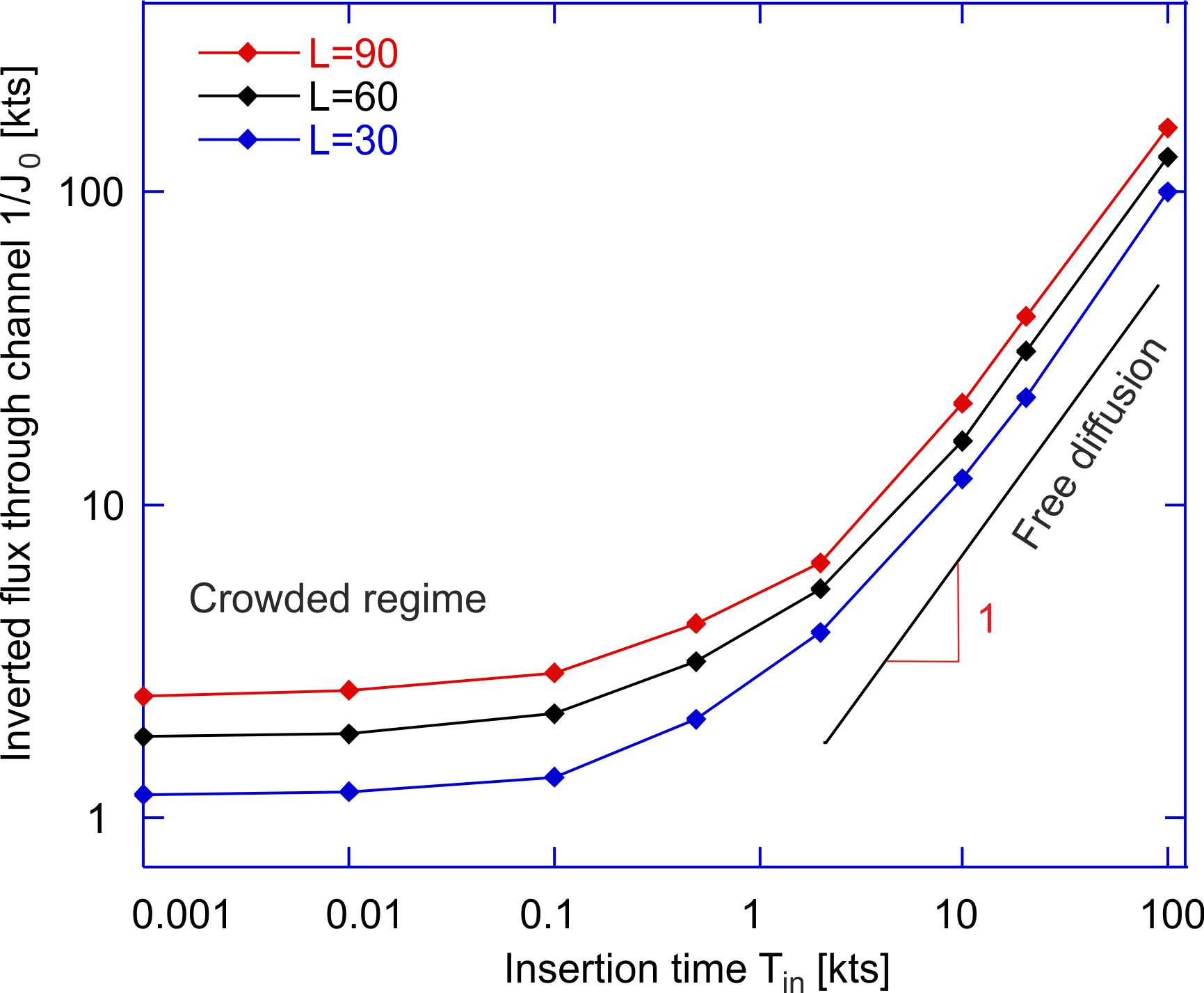}
\caption{The average slope of the time-step versus the cumulative number of particles as a
function of insertion time, $T_{\textrm{in}}$ for three channel lengths. For
high $T_{\textrm{in}}$, $\Delta t/ \Delta N=T_{\textrm{in}} $. The slope
remains constant as a function of $T_{\textrm{in}}$ initially due to the fact
that insertion rate is too high and particles cannot be inserted more often
than $T_{\textrm{in}}\approx 1000$\,ts, the crossover between the constant insertion
rate and the linearly increasing rate at higher $T_{\textrm{in}}$.}\label{fig:flux}
\end{center}
\end{figure}

Since we are interested in the problem of crowding, the most valuable
scenario arises when the insertion rate is high and so particles frequently collide in the
channel. In order to understand how crowding inside the channel affects the
interactions between particles, we considered two scenarios -- an open channel,
as described earlier, in which the particles essentially disappear at the far end (i.e. not able to return back into the channel once they reach $x=L$) -- and a
closed channel with another reflecting wall at $x=L$. The latter channel becomes filled with
particles at a specific rate. Because the length of the channel emerges as an important
parameter, we have again compared channels of
three lengths: $L=30\sigma, 60\sigma$ and $90\sigma$,
fig.\,4. As the time increases, the particles initially fill the channel at a constant rate -- but then their number in channel saturates at a constant value: this is a lower and a highly fluctuating value if the channel is open, and a higher and almost constant value if the channel is closed and the particles pack until there is no more insertion possible. But even the latter value is much less than the channel length: 79 in the channel of 90-particles long, 55 in 60 and 28 in 30.  The difference is due to the effective resistance to insertion of new particles when the density increases sufficiently; note how this difference grows with the length of the channel, which reflects the growing resistance force for new particle insertion.

\begin{figure} 
\begin{center}
  \includegraphics[width=0.37\textwidth]{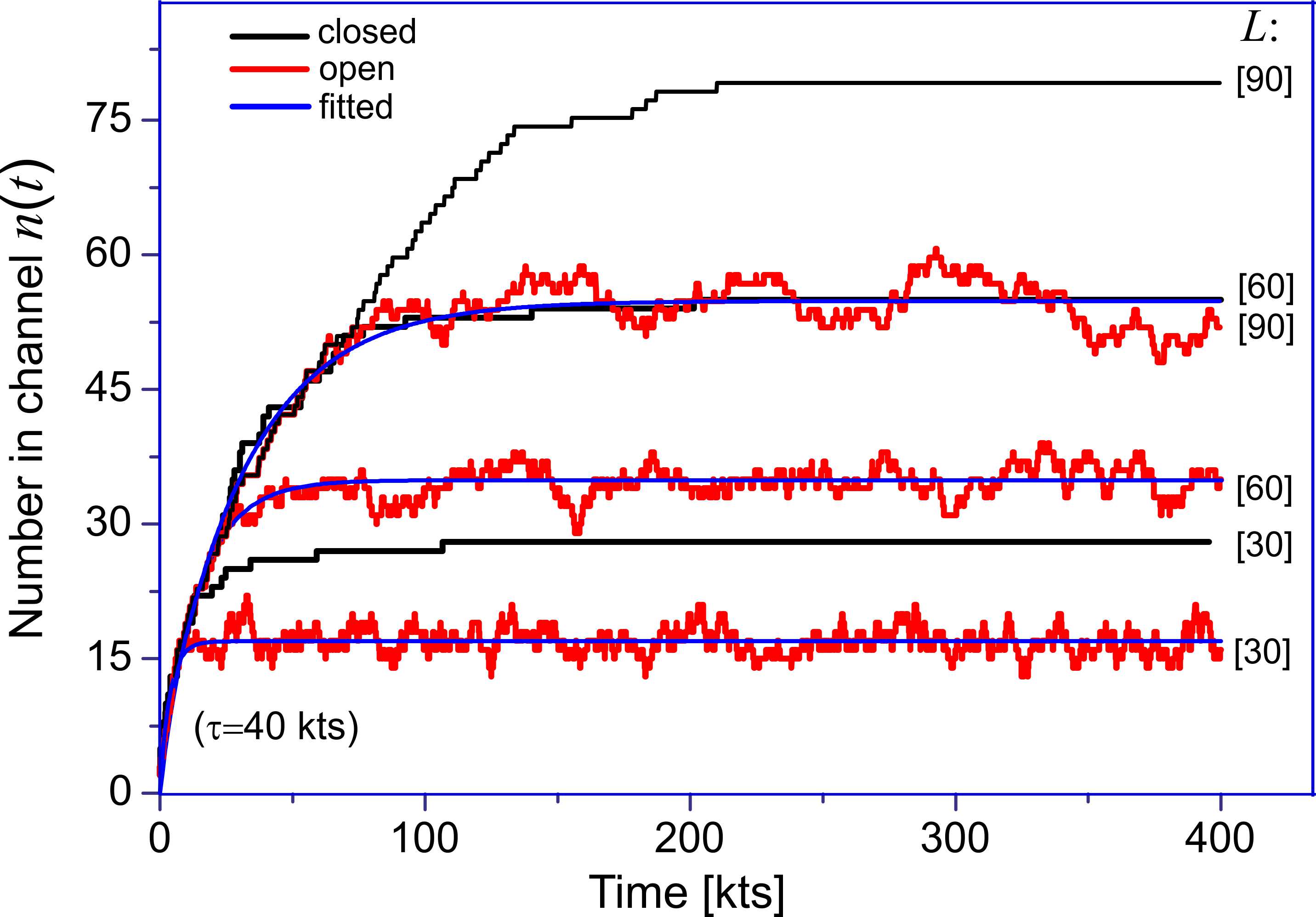}
\caption{The number of particles in open and closed channels of different
lengths: $L= 30\sigma$, $60\sigma$ and $90\sigma$ for the fastest attempted insertion rate ($T_{\textrm{in}} = 0.001$\,kts); the curves are labeled accordingly on the plot. The fluctuations in open channels are caused by the diffusive flux of particles exiting the
channel. Due to a higher overall number of particles in longer channels, the osmotic pressure
inside is higher, which explains the larger gap between $L$ and the mean
plateau values for longer channels.}\label{fig:channels}
\end{center}
\end{figure}

The growth of number of particles in the channel as a function of time was fitted
with the simple exponential function: $n(t) = n_{\textrm{max}} (1- \exp [-t/\tau_\textrm{rel}])$,
where $n_{\textrm{max}}$ is the maximum number of particles in a channel at a given time
(fixed for a channel of given length): the mean plateau value in
fig.\,\ref{fig:channels} gives  $n_{\textrm{max}} = 54$ for $L=90\sigma$, $34$ for $60\sigma$ and $17$ for $30\sigma$.
The gap between the dense-packing and the plateau value in this case (of an open channel) grows approximately linearly with the channel length, in contrast with the closed channel where this gap increases approximately as the square of length.
Remarkably, the characteristic relaxation time $\tau_\textrm{rel}$
after which the dense-packing plateau is achieved was the same in all cases, $\tau_\textrm{rel}
\approx 40$ kts.

\section*{Resistance to transport}

Here we show that the effect of
crowding inside the channel results in an effective potential within
the channel when the density of particles $\rho$ is high. Let us
consider a steady state, at $t > \tau_\textrm{rel}$, when the flux is
constant along the channel, $J_{\textrm{in}}=J_{\textrm{out}}$.
The simple one-dimensional diffusion with the
boundary conditions ($\rho_0$ at $x=0$ and $\rho=0$ at
$r=L$) gives the time-evolving profiles which eventually would reach the steady state
of linear $\rho(x)=[J_\textrm{eq} /D](L-x)$ and constant flux
$J_\textrm{eq}=-D\nabla \rho$. This steady state is only possible if the
density at the entrance $\rho(x=0)=[J_\textrm{eq}/D] L$ remains small,
otherwise crowding becomes relevant in the problem and the simple diffusion
solutions no longer valid.

\begin{figure}[t]
\includegraphics[width=0.37\textwidth]{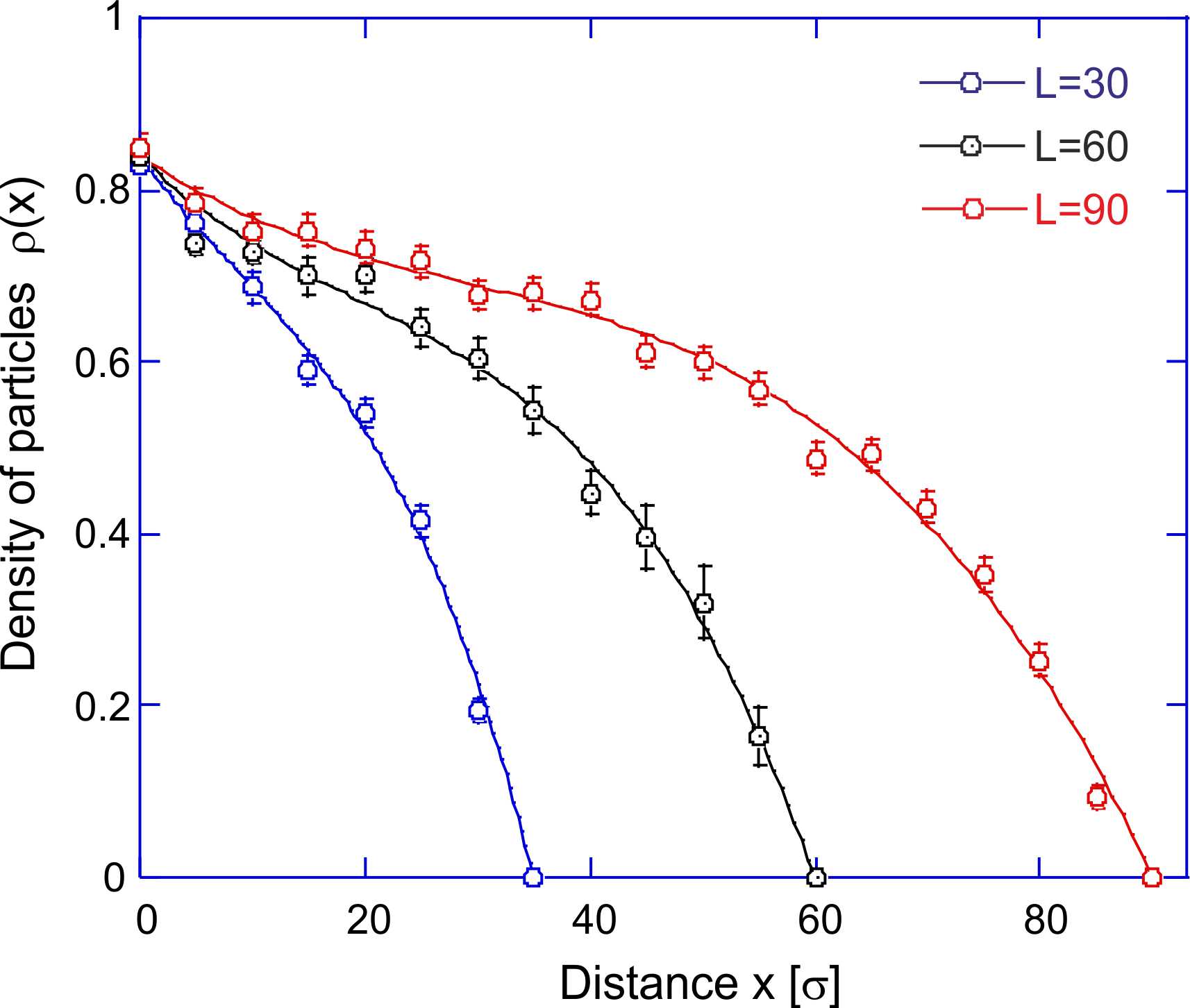}
\caption{The steady-state local density of particles along the
channel of varying lengths (labeled on plot) when it is filled at
the maximum rate $T_{\textrm{in}}=0.001$\,[kts]. Error bars represent 1 standard deviation
based on 80 snapshots at different times during
the equilibrium stage of simulation. The solid lines represent theoretical fitting curves obtained with $\rho=a (L-x) + b(L-x)^{2} + c(L-x)^{3}$ (see eq.\eqref{eq:rho}), where $a=(0.84/L -bL - cL^2)$, and $b=-0.0014, -0.0007,- 0.0004$,  $c= 1.8\cdot 10^{-5}, 6.0\cdot
10^{-6}, 2.2\cdot 10^{-6}$ for $L=30,60,90\sigma$, respectively.}\label{fig:density}
\end{figure}

In order to examine crowding regime in more detail, fig.\,\ref{fig:density} shows the profiles of particle density within the channel at steady-state (in other
words, when the channel is filled at $t > \tau_\textrm{rel}$). The density $\rho (x)$ here is defined
as the number of particles per unit length, measuring the probability density
of finding a particle at a position $x$ in the channel. The deviation from the
linear density profile is prominent.
Generalized diffusion can be used to describe transport at arbitrary particle concentrations.\cite{zaccone_2011,zaccone_2013} The relevant equation reads:
\begin{equation}
\frac{\partial \rho}{\partial t} = \frac{\partial }{\partial x} \left( \left[\frac{D}{k_BT} \frac{\partial \Pi }{\partial \rho} \right] \frac{\partial \rho}{\partial x}  \right) , \label{eq:gendiff}
\end{equation}
where $D_\textrm{coll}=D\, \partial (\beta \Pi)/\partial\rho$, with $\beta = 1/k_BT$, is the collective diffusion coefficient, as defined in nonequilibrium thermodynamics.\cite{dhont} It accounts for deviations from free diffusion through the osmotic pressure $\Pi (\rho)$, which for a non-ideal fluid can be written as a virial expansion in the density $\beta \Pi  =\rho + B_{2}\rho^{2} + B_{3}\rho^{3}$, with $B_{2}$ and $B_{3}$ the second and third virial coefficients, accounting respectively for two-body and three-body interactions between particles in the channel. In bulk 3D systems, higher order coefficients are usually needed in the expansion to properly describe crowding effects at high density,\cite{zaccone_2013} however, we shall see below that in our case of single-file 1D transport terms up to third order are sufficient to capture the phenomenology observed by simulations. This is simply because one particle in front, and one behind, determine the progress of any given particle.
The steady state solution of eq.\eqref{eq:gendiff} is easy to obtain by the chain rule:
\begin{equation}
\Pi  = k_BT  [J_\textrm{eq}/D](L - x).  \label{eq:pi}
\end{equation}
The osmotic pressure has to be a linear function of coordinate along the channel! When the density is low and $\Pi = k_BT \rho$, this reflects the classical steady-state diffusion with the linear density variation $\rho(x)$. At high density we need to invert the dependence $\Pi (\rho )$. Within the third-order virial approximation we obtain
\begin{eqnarray}
\rho  & \approx & \frac{J_\textrm{eq}}{D}(L - x) - B_2 \left( \frac{J_\textrm{eq}}{D} \right)^2 (L-x)^{2}   \label{eq:rho} \\
&& \qquad +(2B_2^2- B_3)  \left( \frac{J_\textrm{eq}}{D} \right)^3 (L-x)^{3} , \nonumber
\end{eqnarray}
where the virial coefficients $B_{2}$ and $B_{3}$ are both positive for repulsive particles used in our simulation. The coefficients $b$ and $c$ used as fitting parameters to successfully reproduce the measured density profiles in fig.\,\ref{fig:density} are explicitly expressed here, and we see that $b<0$ always, while the sign of $c$ depends on the relation between $B_2$ and $B_3$. For our repulsive Lennard-Jones potential the fitting suggests that $2B_2^2-B_3 >0$.

The eq.\eqref{eq:pi} for the osmotic pressure illustrates the first barrier to transport into such a channel. The export apparatus that injects new particles into the channel, at $x=0$, has to exert a force $F =-(1/\rho) [d \Pi/d x]$  to overcome a pressure gradient at the channel entrance: $\Pi (x=0)$ over a distance comparable to the particle size $\sigma$. This ``injection force'' is $F_\textrm{in} \approx  k_BT  [J_\textrm{eq}/D] \, L$, linearly increasing with the channel length if the particle flux is to be maintained constant, as in.\cite{berg_2012} Accordingly, the power required to maintain the constant flux will also be proportional to the channel length: $\dot{W} = k_BT  [J_\textrm{eq}^2 \sigma /D] \, L$.

The effect of crowding on the Brownian resistance force along the channel is described in greater detail in the Appendix. There we show that when the two aditional virial-coefficient terms make the steady-flux density distribution nonlinear, eq.\eqref{eq:rho}, the local force  acting on the particles along the channel acquires an additional contribution due to the collective Brownian effects:
\begin{equation}
F_\textrm{drag}\approx  k_BT\, B_2 \left( \frac{J}{D} \right) k_BT - \frac{1}{2} k_BT \,(B_2^2-B_3) \left( \frac{J }{D} \right)^2L\, .  \label{eq:drag}
\end{equation}
This equation predicts that the (negative) drag force is linear in $L$ for suffiently long channels, and hence to the number of particles in the channel, $n$.

\section*{Discussion}

We have studied ratcheted diffusion transport in nanochannels where translocation is achieved by means of crowding at the channel inlet and depletion at the outlet. This problem is paradigmatic for biomolecule translocation in cellular pores and flagella,\cite{wang_2005} as well as for microfluidics where crowding leads to microchannel clogging.\cite{whitesides,weitz_2006} In a nanochannel, the cargo (protein subunits in a flagellar channel or colloids in a microfluidic channel) has a size comparable to the channel width: the particles move in a single file towards the distal open end, after being injected by an appropriate export machinery from the proximal end. We find many interesting aspects of such motion when the density of particles in the channel becomes high, i.e. their mutual collisions start having a significant effect on the diffusion (crowded regime). Of a particular importance is the criterion of when the crowding regime sets in, which is determined by comparing the injection period $T_\textrm{in} = 1/I_\textrm{in}$ and the time of diffusion along the channel, $\tau_\textrm{diff} = L^2/D$. When the insertion rate is high enough, it takes a certain time to get into the crowded regime; this saturation (clogging) time appears to be independent of the channel length, because it is against the resistance force to the column motion that the clogging occurs.

But the main message and conclusion are expressed at the end of the last section: in order to propel the long column of particles along the channel, maintaining the constant (steady-state) current, the inserting apparatus has to exert an increasing force, and expend an increasing power -- both approximately linearly increasing with the channel length. For long enough channels, the motion will simply stop (channel clogging) until the density slowly drops after gradual ejection of particles from the open end and redistribution of remaining ones along the length. For biological systems such as the flagellin transport, the ATP-powered export apparatus would not be able to increase its power output with the growth of the flagellar length, and instead would have to slow down the rate of insertion to compensate for the growth of resistance. Empirical evidence so far points towards a constant growth rate of flagella, and further investigations are required in the future to clarify the actual mechanism.

In crowded environments, non-specific interactions between molecules lead to deviations from free diffusion. Here we have shown that repulsive interactions between particles lead to an effective drag force, slowing down diffusive transport, which is an emergent cooperative phenomenon. This is an important effect which has to be taken into account in the analysis of many both natural and engineered systems: from biomolecule translocation in membrane pores to the design of microfluidic lab-on-chip devices.

\subsection*{Methods}

We approach the problem of ratcheted diffusion by Brownian dynamics
simulations. Our setup consists of a cylindrical channel into which particles
are added at a certain rate. The force field of the model is rather simple --
there is only hard core repulsion between particles as well as between
particles and the wall of the cylindrical channel, which is represented by the
repulsive part of  the Lennard-Jones potential.  The channel is blocked from
one side, which only allows particles to translocate to the other side of the
channel. As soon as particles reach the end of the channel they disappear from
the system. The dynamics of the process of translocation as well as particle's
positions and forces acting and each particle were recorded at every time-step.
The number of injected - as well as ejected particles was also monitored.
Based on this information, the diffusion profile inside the channel can be
defined as soon as the injection/ejection equilibrium state is reached in the
channel, meaning that at this point, the channel is maximally occupied, and the
rate of injection is equal to the rate of ejection.
New particles are only added when the position at the beginning of the channel
is not occupied by other particles, and therefore even if the particles are set
to be added every time-step, the addition will not happen if another particle
is found less than $\sigma$ distance away from the addition site.
Additionally, we studied how the system behavior changes if the channel is
closed, as opposed to deleting particles when they reach the end of the
channel.

In order to be able to convert the results of simulations, Lennard-Jones
(reduced) units were used. In molecular dynamics algorithms, physical
quantities are expressed as dimensionless units, which results in all physical
quantities being around unity, hence reducing the problems of working with
numbers that are very small or very large, making errors more visible and also
simplifying the output produced and increasing computational efficiency as the
equations of motion become simplified due to the absorption of model-defining
parameters into reduced units. Reduced units also allow one to work on
different problems with a single model. The following reduced units are used
for a Lennard-Jones system: length - $\sigma$, energy - $\epsilon$ and mass
$m$. From these, one can express units of all other physical quantities as follows:
time ($\tau$) as $\tau=\sigma \sqrt{m/\epsilon}$, temperature $k_BT$ in units of
$\epsilon$ and pressure $P$ in units of $\epsilon / \sigma^3$.
The simulation time step was taken as ts$ = 0.01\tau$.

For instance, a `typical' colloid particle of size $\sigma = 1\,\mu$m and mass density $1000$\,kg/m${}^3$ would have a mass $m \approx 4.2\cdot 10^{-15}$kg (staying strictly in SI units here). Taking $k_BT = \epsilon \approx 4.2 \cdot 10^{-21}$J (for room temperature), we obtain the time scale $\tau = 10^{-3}$s. This means that the diffusion constant that we obtained from the fit in fig.\,\ref{fig:tau}, $D \approx 0.0065$ in units [$\sigma^2/\tau$] would have the value $D \approx 6.5 \cdot 10^{-12}$m${}^2$/s, which is almost exactly the magnitude of a widely measured diffusion constant for micron-size sterically stabilized polystyrene particles in water.

\textbf{First passage time.} \ \ 
Here we summarize the standard facts about the first passage problem. This is the classical problem addressing the question: if a diffusing particle starts at, e.g. $x=0$ -- on average how long would it take to reach a given point, $x=L$, for the first time (hence the name of ``first passage'').  Obviously in the stochastic problem one can only have the answer for the average time, and so the aim is to find the probability distribution of such times (and then evaluate the average if required). In our case we remain in the 1-dimensional limit and the generic diffusion equation for the concentration, or the probability to find a particle at $(x,t)$, has the form of continuity relation:
\begin{equation}
\frac{\partial p(x,t)}{\partial t} = - \frac{\partial }{\partial x} J(x,t),
\end{equation}
where $J(x,t)$ is the generalized diffusion flux. For the free diffusion, the Green's function of this equation with $J = -D \nabla p(x,t)$ is the Gaussian
\begin{equation}
G(x,x';t,0)= \frac{1}{\sqrt{4 \pi D t}} e^{-(x-x')^{2}/4Dt}.
\end{equation}
The ``first passage'' condition is implemented by applying the absorbing boundary condition, that is, once the particle first reaches $x=L$, it disappears from the analysis: $p(L,t) =0$. Such a condition is traditionally treated by the method of images, that is, imposing two opposite values -- one starting from $x=0$, the other from $x=2L$:
\begin{equation}
p_\textrm{abs}(x,t)= \frac{1}{\sqrt{4 \pi D t}} \left( e^{-x^{2}/4Dt} - e^{-(2L-x)^{2}/4Dt} \right).  \label{app:p}
\end{equation}
Figure 6 plots several curves of this probability distribution, for $D=1$ and the increasing time (the particles start from $x=0$ and $x=2L$ at $t=0$), to show how the method of images helps to maintain the constraint of $p(L)=0$ at all times.

The survival probability for the particle to remain anywhere between $0$ and $L$, having started at $x=0$ and $t=0$ is obtained by integration: $Q(t) = \int_0^L p_\textrm{abs}(x,t) dx$. The result of this integration is the difference of two error functions:
\begin{equation}
Q(t)= \textrm{erf} \left[ \frac{L}{\sqrt{4 D t}} \right] -  \frac{1}{2} \textrm{erf} \left[ \frac{L}{\sqrt{D t}} \right] .
\end{equation}

Since $x=L$ is an absorbing boundary, the domain of $0-L$ is independent from the domain $L-2L$.  Given the definition of the survival probability, the fraction $-\partial Q(t)/\partial t$ is absorbed between $t$ and $t+dt$. This means that $ f(t) = -\partial Q(t)/\partial t$ is actually the probability density of the time $t$ that takes the particle to reach $x=L$ for the first time, given in the main text as eq.\eqref{Tdiff}.
Figure 7 plots a collection of these distributions for different values of the diffusion constant and fixed $L=1$. In both plots we did not normalize the probability density functions or their parameters, just assuming all values are non-dimensional for qualitative illustration. To find the average value of the first passage time we need to integrate, obtaining the very much expected result:
\begin{equation}
\tau_\textrm{diff} = \int_0^\infty t \, f(t) \, dt = \frac{L^2}{D}.
\end{equation}

\begin{figure} 
\begin{center}
  \includegraphics[width=0.35\textwidth]{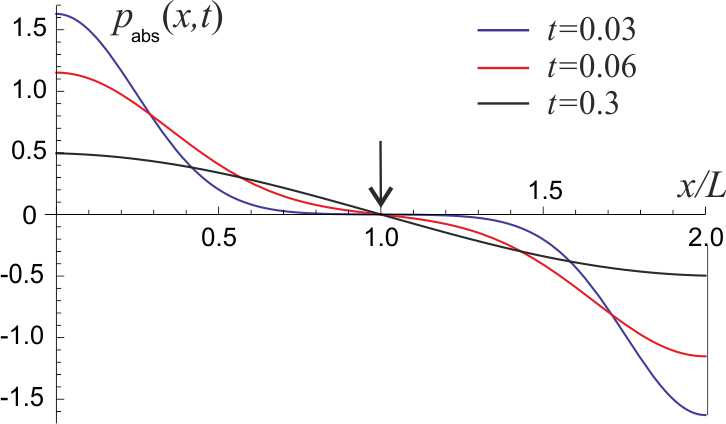}
\caption{ The probability distribution Eq.\eqref{app:p} for particles with the absorbing boundary enforced at $x=L$ by the method of images. }\label{fig:abs}
\end{center}
\end{figure}

\begin{figure} 
\begin{center}
  \includegraphics[width=0.35\textwidth]{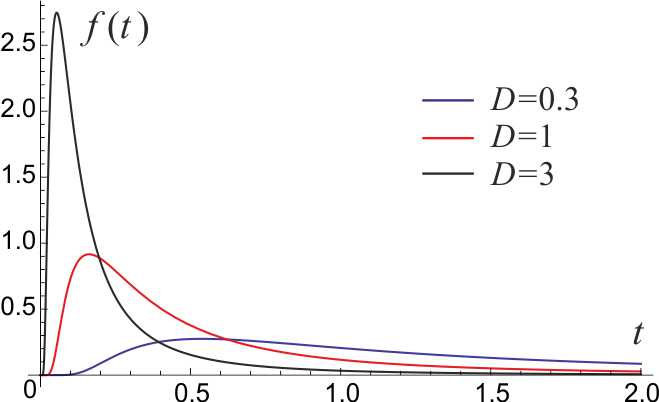}
\caption{ The probability distributions eq.\eqref{Tdiff} of the first passage times in the channel of $0-L$, for several values of the diffusion constant $D$. }\label{fig:pass}
\end{center}
\end{figure}

 \textbf{Forces in crowded channel.} \ \ 
The effect of crowding on the ratcheted transport can be understood via the Brownian force acting on a particle in the system is given in general terms as $F=-(1/\rho)\frac{d \Pi}{d x}$. Since we work in 1-dimensional geometry, with particles unable to overtake each other, only the horizontal projections of all forces matter. We can take as positive all forces parallel to the $x$ axis along the translocation direction (down the density gradient), and with the minus sign we take the resistance forces which oppose the translocation, i.e. drag forces.
Using the chain rule,
\begin{equation}
F=-\frac{1}{\rho}\left( \frac{d \Pi}{d \rho} \right) \frac{d \rho}{d x},   \nonumber
\end{equation}
and employing the virial expansion for $\Pi(\rho)$ and expression for $\rho(x)$ given by the eq.\eqref{eq:rho} of the main text, after minor manipulation we obtain the additional effective force determined by the collective effects in the crowded regime:
\begin{eqnarray}
F_\textrm{crowd} \approx \frac{ k_BT B_2J }{D} -  \frac{k_BT(B_2^2-B_3) J^2}{D^2}(L-x),  \label{eq:fcrowd}
\end{eqnarray}
where the virial coefficients $B_{2}$ and $B_{3}$ were inroduced in the main text.
The last term can be negative, i.e. contribute a net drag force or resistance to diffusive motion.
It is easy to carry out the explicit calculation of the total drag force per particle due to crowding in the channel, by taking the integral over the channel length, with the result given in eq.\eqref{eq:drag}.

It is also interesting to observe that the effect of crowding on diffusive transport can be mapped onto a classical diffusion equation in the effective force-field given by $F_\textrm{crowd}$, in the spirit of an approximation introduced in,\cite{zaccone_2011}
\begin{equation}
\frac{\partial}{\partial x}\left[\left(D\frac{\partial \rho}{\partial x}\right)- \frac{\rho F_\textrm{crowd}}{k_BT} \right]=0
\end{equation}
This is a classical (steady-state) diffusion equation in the force-field $F_\textrm{crowd}$ which takes care of the additional $\rho$-dependent crowding effect.
Hence, the diffusive transport in a crowded environment can be effectively described by means of classical diffusion in the additional force-field due to crowding. The latter can be calculated analytically using \eqref{eq:fcrowd} whenever the inter-particle interaction and the corresponding virial coefficients are known. The solution to this equation is given by the expression for $\rho$ that we presented as eq.\eqref{eq:rho}

\subsubsection*{Acknowledgments}
This work was funded by the Osk. Huttunen Foundation (Finland) and the Ernest
Oppenheimer Fund, Cambridge. The simulations were performed using the Darwin
Supercomputer of the University of Cambridge High Performance Computing Service
(http://www.hpc.cam.ac.uk/), provided by Dell Inc. using Strategic Research
Infrastructure Funding from the Higher Education Funding Council for England.

\end{document}